\documentclass[preprint,showpacs,preprintnumbers,amsmath,amssymb]{revtex4}

\def\brho{{\hbox{\boldmath $\rho$}}}
\def\undersim#1{\setbox9\hbox{${#1}$}{#1}\kern-\wd9\lower
    2.5pt \hbox{\lower\dp9\hbox to \wd9{\hss $_\sim$\hss}}}
\usepackage{amssymb}
\usepackage{amsmath}
\usepackage{graphicx}

\begin{document}

\title{Chaotic parameter $\lambda$ in Hanbury-Brown-Twiss interferometry in an anisotropic
boson gas model}

\author{Jie Liu$^1$}
\author{Peng Ru$^1$}
\author{Wei-Ning Zhang$^{1,\,2}$\footnote{wnzhang@dlut.edu.cn}}
\author{Cheuk-Yin Wong$^3$\footnote{wongc@ornl.gov}}
\affiliation{$^1$School of physics and optoelectronic technology, Dalian University of Technology,
Dalian, Liaoning 116024, China\\
$^2$Department of Physics, Harbin Institute of Technology, Harbin, Heilongjiang 150006, China\\
$^3$Physics Division, Oak Ridge National Laboratory, Oak Ridge, Tennessee 37831, USA}


\begin{abstract}
Using one- and two-body density matrices, we calculate the spatial and momentum distributions, two-particle
Hanbury-Brown Twiss (HBT) correlation functions, and the chaotic parameter $\lambda$ in HBT interferometry
for the systems of boson gas within the harmonic oscillator potentials with anisotropic frequencies in
transverse and longitudinal directions.  The HBT chaotic parameter, which can be obtained by measuring
the correlation functions at zero relative momentum of the particle pair, is related to the degree of
Bose-Einstein condensation and thus the system environment.  We investigate the effects of system temperature, particle number, and the average momentum of the particle pair on the chaotic parameter.
The value of $\lambda$ decreases with the condensed fraction, $f_0$.  It is one for $f_0=0$ and zero for
$f_0=1$.  For a certain $f_0$ between 0 and 1, we find that $\lambda$ increases with the average momentum
of the particle pair and decreases with the particle number of system.  The results of $\lambda$ are
sensitive to the ratio, $\nu=\omega_z/\omega_{\rho}$, of the frequencies in longitudinal and transverse
directions.  They are smaller for larger $\nu$ when $\omega_{\rho}$ is fixed.  In the heavy ion collisions
at the Large Hadron Collider (LHC) energy the large identical pion multiplicity may possibly lead to a
considerable Bose-Einstein condensation.  Its effect on the chaotic parameter in two-pion interferometry
is worth considering in earnest. 
\end{abstract}

\maketitle

\section{Introduction}

Hanbury-Brown-Twiss (HBT) interferometry has been used to study the space-time structure of the
particle-emitting sources in high energy heavy ion collisions \cite{{Gyu79,Won94,Wie99,Wei00,Lis05}}.
The chaotic parameter $\lambda$ in the HBT measurements is introduced phenomenologically to represent
the HBT correlation function at zero relative momentum of two emitted identical bosons.  As is well
known, HBT correlation occurs for chaotic emission and disappears for coherent emission \cite{{Gyu79,
Won94,Wie99,Wei00,Lis05}}.  So, the result of $\lambda$ is related to the chaotic degree of the source,
although there are other effects, such as particle misidentification, long-live resonance decay, final
state Coulomb interaction, non-Gaussian source distribution, and so on, may also lead to $\lambda <1$
for the completely chaotic sources \cite{Wie99,Lis05}.

In Ref. \cite{CsoZim97,ZimCso97}, T. Cs\"{o}rg\H{o} and J. Zim\'{a}nyi investigated the effect of
Bose-Einstein condensation on two-pion interferometry.  They utilized Gaussian formulas describing the
space and momentum distributions of a static non-relativistic boson system, and investigated the influence
of the condensation on pion multiplicity distribution.  In Ref. \cite{Won07}, C. Y. Wong and W. N. Zhang
studied the relationship between the $\lambda$ parameter and the degree of Bose-Einstein condensation
in detail in a boson gas model within a spherical harmonic oscillator potential, which can be analytically
solved in non-relativistic case and be used in atomic physics \cite{Pol96,Nar99,Via06}.  The authors
pointed out that the transition from the condensed coherent phase to the uncondensed chaotic phase
occurs gradually over a large range of temperatures $T$.  A large fraction of pion Bose-Einstein
condensation was estimated in high energy heavy ion collisions on the basis of the spherical boson
gas model.  However, the particle-emitting sources produced in high energy heavy ion collisions are
usually anisotropic in the longitudinal and transverse directions (parallel and perpendicular to the
beam direction).  Investigating the relationship between $\lambda$ and the boson environment for the
anisotropic systems, which is an natural development for the work of Ref. \cite{Won07}, will pave
the way further forward to understand the HBT results in high energy heavy ion collisions.

In this work, we will study the chaotic parameter $\lambda$ in HBT interferometry in a boson gas model
within anisotropic harmonic oscillator potential in transverse and longitudinal directions.  We will
investigate the effects of system environment on the chaotic parameter and estimate the influence of
Bose-Einstein condensation on the values of $\lambda$ in the two-pion HBT measurements in high energy
heavy ion collisions.  Our results indicate that in the anisotropic source model, the chaotic parameter
is not only as a function of the condensed fraction and particle momentum, but also sensitive to the
ratio $\nu=\omega_z/\omega_{\rho}$ of the frequencies in longitudinal and transverse directions.  The
results of $\lambda$ for a larger $\nu$ are smaller than those correspondingly for a smaller $\nu$ and
with the same $\omega_{\rho}$.  The large identical pion multiplicity in the heavy ion collisions at
the Large Hadron Collider (LHC) energy may possibly lead to a considerable Bose-Einstein condensation.
Its effect on the chaotic parameter $\lambda$ in two-pion interferometry is worth considering in earnest.

The rest of this paper is organized as follows.  In Sec. II, we will present the calculations of the
condensed fraction and the formulas of the one- and two-body density matrices for the anisotropic system.
We will study the spatial and momentum density distributions of the systems in Sec. III.  In Sec. IV, we
will calculate the two-particle HBT correlation functions and investigate the influence of Bose-Einstein
condensation on the chaotic parameter $\lambda$ in HBT interferometry.  We will investigate the influence
of Bose-Einstein condensation on the $\lambda$ value in the two-pion HBT measurements in high energy heavy
ion collisions.  Finally, the summary and discussion will be given in Sec. V.

\section{The boson gas in anisotropic harmonic oscillator potential}

Following the previous works \cite{Pol96,Nar99,Via06,Won07}, we consider a model of the ideal boson gas
held together in a harmonic oscillator potential, $V(\textbf{\emph{r}})$, that arises either externally
(in atomic physics) or from bosons¡¯ own mean fields (in high-energy heavy-ion collisions).  We use
$\hbar\,\omega_{\rho}$ and $\hbar\,\omega_z$ to measure the strengths of the potential in transverse and
longitudinal directions,
\begin{equation}
\label{Vrz}
V(\textbf{\emph{r}})=\frac{1}{2}\,m\,\omega_{\rho}^2\, \rho^2 +\frac{1}{2}\,m\,\omega_z^2\, z^2
=\frac{1}{2}\,\hbar\,\omega_{\rho} \bigg(\frac{\rho}{a_{\rho}}\bigg)^2
+\frac{1}{2}\,\hbar\,\omega_z \bigg(\frac{z}{a_z}\bigg)^2,
\end{equation}
where, $\rho=\sqrt{x^2+y^2}$ and $z$ are transverse and longitudinal coordinates, $m$ is mass of a boson,
and $a_{\rho}=\sqrt{\hbar/m\omega_{\rho}}$ and $a_z=\sqrt{\hbar/m\omega_z}$ are two length parameters of
the system in the transverse and longitudinal directions, which are related to $\omega_{\rho}$ and $\omega_z$,
respectively.

For the ideal boson gas, the system energy is simply the summation of all individual bosons, and the energy
levels of a boson in the anisotropic harmonic oscillator potential are
\begin{eqnarray}
\label{enen}
\varepsilon_n&=&(n_x+\frac{1}{2})\hbar\,\omega_{\rho}+(n_y+\frac{1}{2})\hbar\,\omega_{\rho}+
(n_z+\frac{1}{2})\hbar\,\omega_z\cr
&=&(n_{\rho}+1) \hbar\,\omega_{\rho}+(n_z+\frac{1}{2}) \hbar\,\omega_z,
~~~~n_{\rho}=n_x+n_y,
\end{eqnarray}
where, $n=0, 1, 2, \cdots$ are the indexes of the energy levels $\varepsilon_0 < \varepsilon_1
< \varepsilon_2 < \cdots$, and $n_i=0,1,2,...\,(i=x,y,z)$.  Introducing $\tilde{\varepsilon}_{n_{\rho}}
= n_{\rho} \hbar\,\omega_{\rho}$ and $\tilde{\varepsilon}_{n_z}=n_z \hbar\,\omega_z$, we have
\begin{equation}
\varepsilon_n=\tilde{\varepsilon}_{n_{\rho}}+\tilde{\varepsilon}_{n_z}+\varepsilon_0,
~~~~ n_{\rho}, n_z=0,1,2,...\,,
\end{equation}
where $\varepsilon_0=\hbar\,\omega_{\rho}+\frac{1}{2}\hbar\,\omega_z$ is the lowest energy level for
the ground state $n=0$ ($n_{\rho}=n_z=0$).  For a certain $n_{\rho}$, the degeneracy of the energy
level $\tilde{\varepsilon}_{n_{\rho}}$ of the transverse two-dimension oscillator is $(n_{\rho}+1)$.
And, the degeneracy of the energy level $\tilde{\varepsilon}_{n_z}$ of the longitudinal one-dimension
oscillator is one for each $n_z$ value.  Each energy level $\varepsilon_n$ corresponds to a certain
$n_{\rho}$ and $n_z$.  The degeneracy of $\varepsilon_n$ is $g_n=(n_{\rho}+1)$.  For instance, the
energy levels and their degeneracies for the case of $\omega_{\rho}<\omega_z<2\omega_{\rho}$ are:
\begin{eqnarray}
\varepsilon_0&=&\hbar\,\omega_{\rho}+\hbar\,\omega_z/2,~~~~n_{\rho}=(n_x+n_y)=0,~~~n_z=0,~~~g_0=1,\cr
\varepsilon_1&=&2\hbar\,\omega_{\rho}+\hbar\,\omega_z/2,~~~n_{\rho}=(n_x+n_y)=1,~~~n_z=0,~~~g_0=2,\cr
\varepsilon_2&=&\hbar\,\omega_{\rho}+3\hbar\,\omega_z/2,~~~n_{\rho}=(n_x+n_y)=0,~~~n_z=1,~~~g_0=1,\cr
\varepsilon_3&=&3\hbar\,\omega_{\rho}+\hbar\,\omega_z/2,~~~n_{\rho}=(n_x+n_y)=2,~~~n_z=0,~~~g_0=3,\cr
&&\cdots~~~~~~~~\cdots~~~~~~~~\cdots\nonumber
\end{eqnarray}

\subsection{Condensed fraction}

For the identical boson gas with a fixed number of particles, $N$, and at a given temperature
$T=1/\beta$, we have
\begin{equation}
\label{N0T}
N=N_0+N_T,
\end{equation}
where, $N_0$ is the number of condensate particles in $n=0$ state,
\begin{equation}
\label{N0}
N_0=\frac{\mathcal Z}{1-\mathcal Z},
\end{equation}
$N_T$ is the number of the particles in $n>0$ states,
\begin{equation}
\label{NTn}
N_T=\sum_{n>0}^{\infty}\frac{g_n \mathcal Z\,e^{-\beta(\tilde{\epsilon}_{n_{\rho}}+\tilde{\epsilon}_{n_z})
}}{1-\mathcal{Z}\,e^{-\beta(\tilde{\epsilon}_{n_{\rho}}+\tilde{\epsilon}_{n_z})}},
\end{equation}
and $\mathcal Z$ is the fugacity parameter which includes the factor for the lowest energy $\varepsilon_0$
\cite{Nar99}.  Because $N_0 \ge 0$, the values of $\mathcal Z$ are between zero and unity.  When the
temperature of the gas is lowered below the critical temperature $T_{\rm c}$, condensation of the boson gas
occurs.  In this case, $N_0 \sim N$ and $\mathcal Z \sim N/(N+1)$.  In Eq. (\ref{NTn}), the denominator can
be expanded as
\begin{equation}
\frac{1}{1-\mathcal{Z}\,e^{-\beta(\tilde{\epsilon}_{n_{\rho}}+\tilde{\epsilon}_{n_z})}}=\sum_{k=0}^{\infty}
\mathcal{Z}^k\,e^{-k\beta(\tilde{\epsilon}_{n_{\rho}}+\tilde{\epsilon}_{n_z})}.\nonumber
\end{equation}
And, Eq. (\ref{NTn}) can be written as
\begin{eqnarray}
N_{T}&=&\sum_{n_{\rho}+n_z>0}^{\infty} (n_{\rho}+1) \mathcal{Z}\,e^{-\beta\tilde{\epsilon}_{n_{\rho}}}\, e^{-\beta\tilde{\epsilon}_{n_z}} \sum_{k=0}^{\infty} \mathcal{Z}^k\,e^{-k\beta\tilde{\epsilon}_{n_{\rho}}}\, e^{-k\beta\tilde{\epsilon}_{n_z}}\cr
&=&\sum_{n_{\rho}+n_z>0}^{\infty} (n_{\rho}+1) \sum_{k=1}^{\infty} \mathcal{Z}^k\, e^{-k\beta\tilde{\epsilon}_{n_{\rho}}}\, e^{-k\beta\tilde{\epsilon}_{n_z}}\cr
&=&\sum_{k=1}^{\infty} \mathcal{Z}^k \bigg[\sum_{n_{\rho}=0}^{\infty} (n_{\rho}+1) e^{-k\beta n_{\rho}
\hbar \omega_{\rho}} \sum_{n_z=0}^{\infty} e^{-k\beta n_z \hbar \omega_z} -1 \bigg ]\cr
&=&\sum_{k=1}^{\infty} \mathcal{Z}^k \bigg[\sum_{n_x=0}^{\infty} e^{-k\beta n_x \hbar\omega_{\rho}} \sum_{n_y=0}^{\infty} e^{-k\beta n_y \hbar\omega_{\rho}} \sum_{n_z=0}^{\infty} e^{-k\beta n_z
\hbar\omega_z} -1 \bigg ],\nonumber
\end{eqnarray}
where the last term 1 corresponds $n_{\rho}=n_x+n_y=0$ and $n_z=0$.  So, Eq. (\ref{NTn}) can be written
as
\begin{eqnarray}
\label{NTk}
N_{T}&=&\sum^{\infty}_{k=1} \mathcal Z^k \bigg [\frac{1}{(1-e^{-k\beta\hbar\omega_{\rho}})^2} \cdot \frac{1}{1
-e^{-k\beta\hbar\omega_z}}-1 \bigg]\cr
&=&\sum^{\infty}_{k=1} \mathcal Z^k \bigg [\frac{e^{-k\beta\hbar\omega_{\rho}}(2 - e^{-k\beta\hbar
\omega_{\rho}} + e^{-k\beta\hbar\omega_{\rho}}e^{-k\beta\hbar\omega_z} -2e^{-k\beta\hbar
\omega_z}) + e^{-k\beta\hbar\omega_z}}{(1-e^{-k\beta\hbar\omega_{\rho}})^2 (1-e^{-k\beta\hbar
\omega_z})} \bigg].
\end{eqnarray}

\begin{figure}[tbp]
\includegraphics[width=0.52\columnwidth]{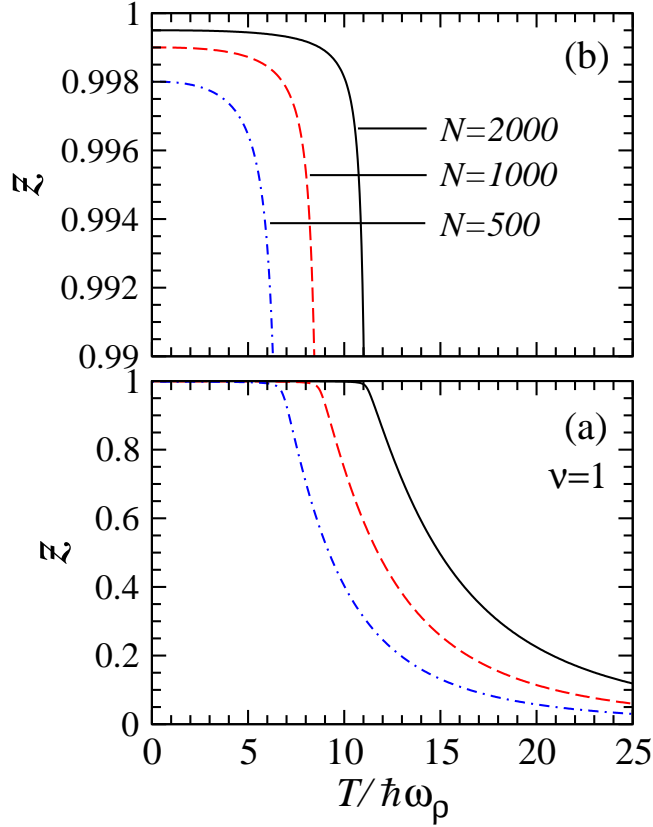}
\caption{(Color online) (a) The fugacity parameter $\mathcal Z$ as a function of $T/\hbar\omega_{\rho}$
for different boson numbers $N$ and $\nu=(\omega_z/\omega_{\rho})=1$.  (b) An expanded view in the
$\mathcal Z\sim1$ region. } \label{zTN}
\end{figure}

\begin{figure}[tbp]
\includegraphics[width=0.52\columnwidth]{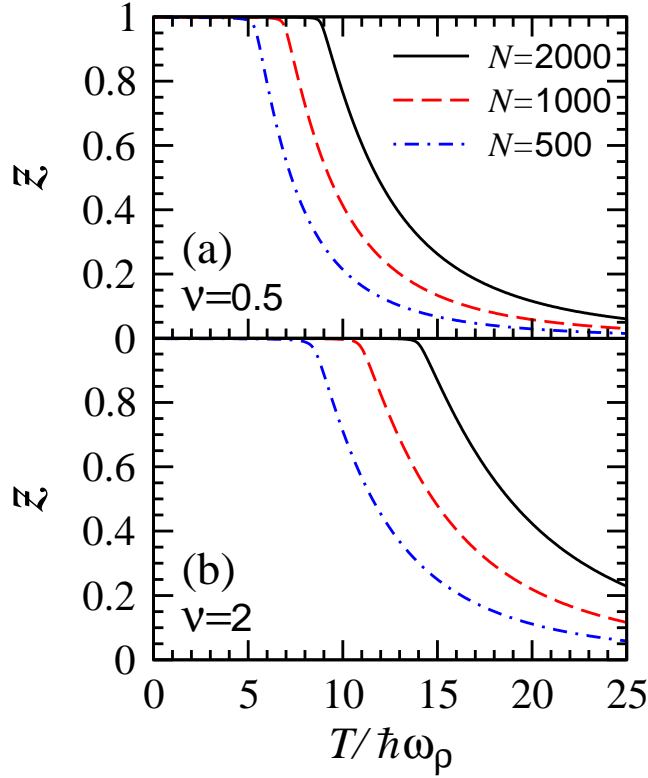}
\caption{(Color online) The fugacity parameter $\mathcal Z$ as a function of $T/\hbar\omega_{\rho}$ and
$N$ for (a) $\nu=(\omega_z/\omega_{\rho})=0.5$ and (b) $\nu=(\omega_z/\omega_{\rho})=2$. } \label{zTN2}
\end{figure}

From Eqs. (\ref{N0T}), (\ref{N0}), and (\ref{NTk}), one can determine the fugacity parameter $\mathcal Z$
as a function of $T$ and $N$ numerically \cite{Won07}.  In Fig. \ref{zTN}, we show the solution of
$\mathcal Z$ for different temperatures $T/\hbar\,\omega_{\rho}$ and boson numbers $N$.  Here $\nu$ is
the parameter of the ratio of $\omega_z$ to $\omega_{\rho}$, $\nu=\omega_z/\omega_{\rho}$.  To get a
better view of the $\mathcal Z$ values, we show an expanded view of Fig. \ref{zTN}(a) in the $\mathcal Z
\sim1$ region in Fig. \ref{zTN}(b).  One can see that the fugacity parameter $\mathcal Z$ is close to
unity in the highly condensation region at low temperatures.  For a given $N$, as the temperature increases
from zero, the fugacity $\mathcal Z$ decreases very slowly in the form of a plateau until the critical
temperature $T_c$ is reached, and it decreases much rapidly thereafter.  The transition from the condensed
phase to the uncondensed phase occurs over a large range of temperatures.  The greater the number of bosons
$N$, the greater is the plateau region, as shown in Fig. \ref{zTN}(b).  The width of the $T/\hbar\,\omega$
plateau for $N=2000$ is about 11.  And, the widths of the plateaus for $N=500$ is about a half of that
for $N=2000$.  Figures \ref{zTN2}(a) and \ref{zTN2}(b) show $\mathcal Z$ as a function of $T/\hbar\,
\omega_{\rho}$ and $N$ for $\nu=0.5$ and 2, respectively.  One can see that the width of the $\mathcal Z$
plateau is sensitive to the ratio $\nu=\omega_z/\omega_{\rho}$.  For a fixed $N$, the greater the ratio
$\nu$, the greater is the plateau.

\begin{figure}[tbp]
\includegraphics[width=0.43\columnwidth]{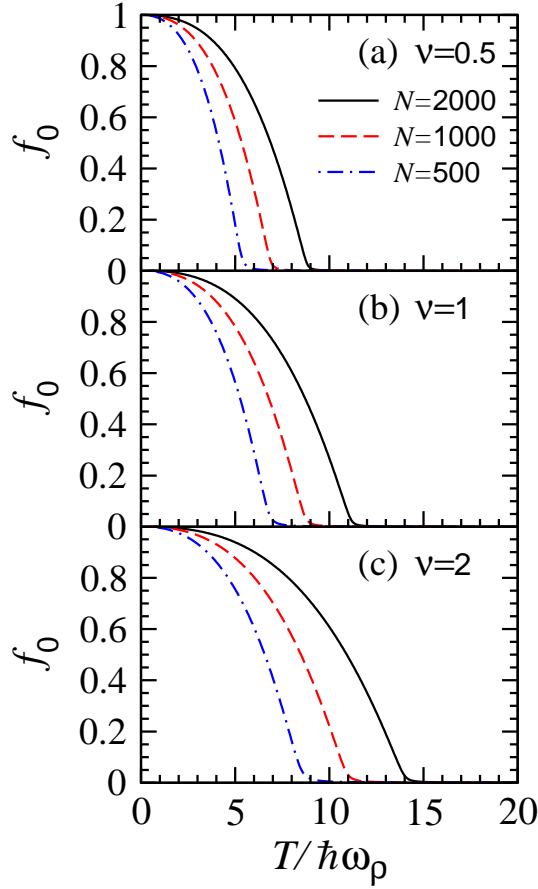}
\caption{(Color online) The condensed fraction $f_{0}$ as a function of $T/\hbar\omega_{\rho}$, $N$,
and $\nu$. }
\label{f0}
\end{figure}

In the transition region $T\,\undersim<\,T_c$, the fugacity parameter $\mathcal Z\sim 1$.  By setting
$\mathcal Z$ to unity in Eq. (\ref{NTk}), and considering $\hbar\,\omega_{\rho}/T \ll 1$ \cite{Won07},
we have
\begin{eqnarray}
N_{T}\sim \sum^{\infty}_{k=1} \bigg [\frac{1}{(k\beta\hbar\omega_{\rho})^2 (k\beta\hbar\omega_z)}\bigg]
=\bigg(\frac{T}{\hbar\omega_{\rho}}\bigg)^2 \bigg(\frac{T}{\hbar\omega_{\rho}\nu}\bigg) \sum^{\infty}_{k=1}\frac{1}{k^3}=\bigg(\frac{T}{\hbar\omega_{\rho}}\bigg)^3\frac{1}{\nu}\,\zeta(3).
\nonumber
\end{eqnarray}
Considering $N_T\sim N$ at $T_{\rm c}$, we obtain
\begin{equation}
\bigg(\frac{T_{\rm c,\,approx}}{\hbar\,\omega_{\rho}}\bigg) \sim \bigg[\frac{N}{\zeta(3)}\,\nu
\bigg]^{1/3} = \bigg[\frac{N}{1.202}\,\nu\bigg]^{1/3}.
\end{equation}
For $\nu=1$, it becomes the result of three-dimension isotropic harmonic oscillator \cite{Won07,Pol96}.
For $N=2000$ the approximated critical temperatures for $\nu=0.5$, 1, and 2 are $T_{\rm c,\,approx}
=9.40$, 11.85, and 14.93 $\hbar\,\omega_{\rho}$, respectively.  They are approximately consistent with
the results in Figs. \ref{zTN} and \ref{zTN2}.

After obtaining the value of the solution $\mathcal Z$, one can specify the condensed fraction $f_0$ and
the uncondensed fraction $f_T$ by
\begin{equation}
f_0=\frac{N_{0}}{N}~~~~~{\rm and}~~~~~f_T=\frac{N_T}{N}=1-f_0.
\end{equation}
Figures \ref{f0}(a), \ref{f0}(b), and \ref{f0}(c) show the condensed fraction $f_0$ as a function of
$T/\hbar\,\omega_{\rho}$ and $N$ for the ratios $\nu=$0.5, 1, and 2, respectively.  The values of
$f_0$ are unity at $T=0$, corresponding to a completely coherent boson system at $T=0$.  The fraction
decreases slowly as the temperature increases, and the rate of decrease is small at low temperatures.
For a fixed $\nu$, the greater the number of bosons $N$, the larger is the range of temperatures in
which the boson system contains a substantial fraction of condensation.  However, for a fixed $N$
and $T$, the condensed fraction increases with $\nu$.  One can see that for the systems with 2000
identical bosons, substantial fractions of condensation occur up to $T/\hbar\,\omega \sim$9, 11, and
14 for $\nu=$0.5, 1, and 2, respectively.

\subsection{One-body and two-body density matrices}

In order to evaluate the system spatial and momentum distributions and two-particle momentum correlation
functions, we need write down the one-body and two-body density matrices in configuration and momentum
spaces.  The one-body density matrix in configuration space for the boson gas with fixed bosons number
$N$ and in three-dimension harmonic oscillator potential is
\begin{eqnarray}
\label{G1r1}
G^{(1)}_{\omega_x\omega_y\omega_z}(\textbf{\emph{r}}_1,\textbf{\emph{r}}_2)
&=&\sum_{n_x,n_y,n_z} \frac{\mathcal Z\,e^{-\beta n_x\hbar\omega_x} e^{-\beta n_y \hbar\omega_y} e^{-\beta
n_z \hbar\omega_z}}{1-\mathcal Z\,e^{-\beta n_x \hbar\omega_x} e^{-\beta n_y \hbar\omega_y} e^{-\beta n_z
\hbar\omega_z}} U_{n_x}(x_1,\omega_x)\cr\cr
&&\hspace*{-20mm}\times\,U_{n_y}(y_1,\omega_y)U_{n_z}(z_1,\omega_z)U_{n_x}^*(x_2,\omega_x)U_{n_y}^*
(y_2,\omega_y)U_{n_z}^*(z_2,\omega_z)\cr\cr
&&\hspace*{-20mm}=\,\sum_{n_x,n_y,n_z}\sum_{k=1}^{\infty} \mathcal Z^k e^{-\beta k n_x \hbar\omega_x}
e^{-\beta k n_y \hbar\omega_y} e^{-\beta k n_z \hbar\omega_z}U_{n_x}(x_1,\omega_x)\cr\cr
&&\hspace*{-20mm}\times\,U_{n_y}(y_1,\omega_y)U_{n_z,}(z_1,\omega_z)U_{n_x}^*(x_2,\omega_x)
U_{n_y}^*(y_2,\omega_y)U_{n_z}^*(z_2,\omega_z),
\end{eqnarray}
where
\begin{eqnarray}
\label{wfx}
U_{n_x}(x,\omega_x)=\bigg(\frac{1}{2^{n_x}\, n_x!}\bigg)^{1/2} \bigg(\frac{m\omega_x}{\pi\hbar}
\bigg)^{1/4}\exp{\bigg(-\frac{m\omega_x x^2}{2\hbar}\bigg)} H_{n_x}\bigg(\sqrt{\frac{m\omega_x}{\hbar}}
x\bigg).
\end{eqnarray}
Using the property of Hermitian function,
\begin{equation}
\exp{[-(\xi^2+\eta^2)]}\sum_{n=0}^{\infty}\bigg(\frac{\zeta^n}{2^n\,n!}\bigg) H_n(\xi) H_n(\eta)=
\frac{1}{\sqrt{1-\zeta^2}}\exp\bigg(-\frac{\xi^2+\eta^2-2\xi\zeta\eta}{1-\zeta^2}\bigg),
\end{equation}
and setting $\omega_x=\omega_y=\omega_{\rho}$ in Eq. (\ref{G1r1}), we have
\begin{equation}
\label{G1s}
G^{(1)}_{\omega_{\rho}\omega_z}(\textbf{\emph{r}}_1,\textbf{\emph{r}}_2)=\sum_{k=1}^{\infty}
\mathcal Z^k \tilde{G}_0(\textbf{\emph{r}}_1,\textbf{\emph{r}}_2;\beta k\hbar\, \omega_{\rho}, \beta
k\hbar\,\omega_z),
\end{equation}
where
\begin{eqnarray}
\label{G0s}
&&\tilde{G}_0(\textbf{\emph{r}}_1,\textbf{\emph{r}}_2; \tau_{\rho},\tau_z)\cr\cr
&&=\frac{1}{\pi a_{\rho}^2(1-e^{-2\tau_{\rho}})} \exp\bigg[-\frac{1}{a_{\rho}^2}\frac{(\brho_1^2
+\brho_2^2)(\cosh \tau_{\rho}-1) +(\brho_1 -\brho_2)^2}{2\sinh \tau_{\rho}}\bigg]\cr\cr
&&\times \bigg[\frac{1}{\pi a_z^2(1-e^{-2\tau_z})}\bigg]^{\frac{1}{2}}\exp\bigg[-
\frac{1}{a_z^2}\frac{(z_1^2+z_2^2)(\cosh \tau_z-1)+(z_1 -z_2)^2}{2\sinh \tau_z} \bigg],
\end{eqnarray}
where $\tau_{\rho}=\beta k\hbar\,\omega_{\rho}$ and $\tau_z=\beta k\hbar\,\omega_z$.

Using the wave function of ground state,
\begin{equation}
\label{wfs}
U_0(\textbf{\emph{r}}) =\bigg(\frac{1}{\pi a_{\rho}^2}\bigg)^{\!\frac{1}{2}}
\bigg(\frac{1}{\pi a_z^2}\bigg)^{\!\frac{1}{4}} \exp\bigg(\!\!-\frac{\brho^2}{2a_{\rho}^2}
-\frac{z^2}{2 a_z^2}\bigg),
\end{equation}
we may further write $\tilde{G}_0(\textbf{\emph{r}}_1,\textbf{\emph{r}}_2; \tau_{\rho},\tau_z)$
as
\begin{equation}
\tilde{G}_0(\textbf{\emph{r}}_1,\textbf{\emph{r}}_2;\tau_{\rho},\tau_z)=U_0^*(\textbf{\emph{r}}_1)
U_0(\textbf{\emph{r}}_2)\,\tilde{g}_0 (\textbf{\emph{r}}_1,\textbf{\emph{r}}_2;\tau_{\rho},\tau_z),
\end{equation}
where
\begin{eqnarray}
\label{sG0}
&&\tilde{g}_0(\textbf{\emph{r}}_1,\textbf{\emph{r}}_2; \tau_{\rho},\tau_z)\cr\cr
&&=\frac{1}{(1-e^{-2\tau_{\rho}})} \exp\bigg[ -\frac{1}{a_{\rho}^2}
\frac{(\brho_1^2 +\brho_2^2)(\cosh\tau_{\rho} -1 -\sinh\tau_{\rho})+(\brho_1 -\brho_2)^2}{2\sinh
\tau_{\rho}}\bigg]\cr\cr
&&\times \frac{1}{(1-e^{-2\tau_z})^{1/2}}\exp\bigg[-\frac{1}{a_z^2}
\frac{(z_1^2+ z_2^2)(\cosh\tau_z-1-\sinh\tau_z)+(z_1-z_2)^2}{2\sinh\tau_z}\bigg].
\end{eqnarray}\\
Then, we have
\begin{equation}
\label{G1A}
G^{(1)}_{\omega_{\rho}\omega_z}(\textbf{\emph{r}}_1,\textbf{\emph{r}}_2)
=U_0^*(\textbf{\emph{r}}_1)U_0(\textbf{\emph{r}}_2)A_{\omega_{\rho}\omega_z}(\textbf{\emph{r}}_1,
\textbf{\emph{r}}_2),
\end{equation}
where
\begin{equation}
\label{Ap12}
A_{\omega_{\rho}\omega_z}(\textbf{\emph{r}}_1,\textbf{\emph{r}}_2)=\sum_{k=1}^{\infty}\mathcal Z^k\,
\tilde{g}_0(\textbf{\emph{r}}_1,\textbf{\emph{r}}_2;\beta k\hbar\,\omega_{\rho},\beta k\hbar\,
\omega_z).
\end{equation}

The one-body density matrix in momentum space can be readily obtained from the equations of
$G^{(1)}_{\omega_{\rho}\omega_z}(\textbf{\emph{r}}_1,\textbf{\emph{r}}_2)$ by using the symmetry
between $\textbf{\emph{r}}/a$ and $\textbf{\emph{p}}a/\hbar$ in a harmonic oscillator potential,
and we get
\begin{equation}
\label{G1Am}
G^{(1)}_{\omega_{\rho}\omega_z}(\textbf{\emph{p}}_1,\textbf{\emph{p}}_2)
=U_0^*(\textbf{\emph{p}}_1)U_0(\textbf{\emph{p}}_2)A_{\omega_{\rho}\omega_z}(\textbf{\emph{p}}_1,
\textbf{\emph{p}}_2),
\end{equation}
where
\begin{equation}
\label{wfm}
U_0(\textbf{\emph{p}}) =\bigg(\frac{a_{\rho}^2}{\pi\hbar^2}\bigg)^{\!\frac{1}{2}}
\bigg(\frac{a_z^2}{\pi\hbar^2}\bigg)^{\!\frac{1}{4}} \exp\bigg(\!\!-\frac{a_{\rho}^2}{\hbar^2}
\frac{\textbf{\emph{p}}_{\!\rho}^2}{2} -\frac{a_z^2}{\hbar^2}\frac{p_z^2}{2}\bigg)
\end{equation}
is the ground state wave function in momentum space, and
\begin{equation}
\label{Ap12}
A_{\omega_{\rho}\omega_z}(\textbf{\emph{p}}_1,\textbf{\emph{p}}_2)=\sum_{k=1}^{\infty}\mathcal Z^k\,
\tilde{g}_0(\textbf{\emph{p}}_1,\textbf{\emph{p}}_2;\beta k\hbar\,\omega_{\rho},\beta k\hbar\,
\omega_z),
\vspace*{-8mm}
\end{equation}
where
\begin{eqnarray}
\label{sG0}
&&\tilde{g}_0(\textbf{\emph{p}}_1,\textbf{\emph{p}}_2; \tau_{\rho},\tau_z)\cr\cr
&&=\frac{1}{(1-e^{-2\tau_{\rho}})} \exp\bigg[ -\frac{a_{\rho}^2}{\hbar^2}
\frac{(\textbf{\emph{p}}_{\!1\rho}^2 +\textbf{\emph{p}}_{2\rho}^2)(\cosh\tau_{\rho} -1
-\sinh\tau_{\rho})+(\textbf{\emph{p}}_{\!1\rho} -\textbf{\emph{p}}_{2\rho} )^2}{2\sinh\tau_{\rho}}
\bigg]\cr\cr
&&\times \frac{1}{(1-e^{-2\tau_z})^{1/2}}\exp\bigg[-\frac{a_z^2}
{\hbar^2} \frac{(p_{1z}^2+p_{2z}^2)(\cosh\tau_z-1-\sinh\tau_z)+(p_{1z}-p_{2z})^2}{2\sinh\tau_z}\bigg].
\end{eqnarray}

In the limit of a large number of particles, $N_0^2\gg N_0$, and the two-body density matrix in
momentum space can be written as \cite{Pol96,Nar99,Won07}
\begin{eqnarray}
\label{G2m}
G^{(2)}_{\omega_{\rho}\omega_z}(\textbf{\emph{p}}_1,\textbf{\emph{p}}_2;\textbf{\emph{p}}_1,
\textbf{\emph{p}}_2)&=&G^{(1)}_{\omega_{\rho}\omega_z}(\textbf{\emph{p}}_1,\textbf{\emph{p}}_1)\,
G^{(1)}_{\omega_{\rho}\omega_z}(\textbf{\emph{p}}_2,\textbf{\emph{p}}_2) +|\,G^{(1)}_{\omega_{\rho}\omega_z}
(\textbf{\emph{p}}_1,\textbf{\emph{p}}_2)|^2\cr\cr
&-&N_0^2|U_0(\textbf{\emph{p}}_1)|^2 |U_0(\textbf{\emph{p}}_2)|^2.
\end{eqnarray}
From Eqs. (\ref{N0}) and (\ref{G1Am}), we have
\begin{eqnarray}
\label{G2m2}
G^{(2)}_{\omega_{\rho}\omega_z}(\textbf{\emph{p}}_1,\textbf{\emph{p}}_2;\textbf{\emph{p}}_1,
\textbf{\emph{p}}_2)&=& |U_0(\textbf{\emph{p}}_1)|^2|U_0(\textbf{\emph{p}}_2)|^2 \big[ A_{\omega_{\rho}\omega_z}
(\textbf{\emph{p}}_1,\textbf{\emph{p}}_1)\,
A_{\omega_{\rho}\omega_z}(\textbf{\emph{p}}_2,\textbf{\emph{p}}_2)\cr\cr
&&+|\,A_{\omega_{\rho}\omega_z} (\textbf{\emph{p}}_1,\textbf{\emph{p}}_2)|^2 -[\mathcal Z/(1-
\mathcal Z)]^2\,\big].
\end{eqnarray}

In numerical calculations, it is useful to rewrite Eq. (\ref{Ap12}) as
\begin{eqnarray}
\label{A2}
A_{\omega_{\rho}\omega_z}(\textbf{\emph{p}}_1,\textbf{\emph{p}}_2)=\frac{\mathcal Z}{1-\mathcal Z}
+\sum_{k=1}^{\infty} \mathcal Z^k \Big[\tilde{g}_0(\textbf{\emph{p}}_1,\textbf{\emph{p}}_2;
\beta k\hbar\,\omega_{\rho},\beta k\hbar\,\omega_z)-1\Big],
\end{eqnarray}
for a rapid convergence of the summation \cite{Won07}.  For low temperatures, $[\tilde{g}_0(
\textbf{\emph{p}}_1,\textbf{\emph{p}}_2;\tau_{\rho},\tau_z)-1]$ is small and a small number of
terms in $k$ will suffice.  For high temperatures above critical temperature, $\mathcal Z\ll 1$, and a
small number of terms in $k$ will also suffice because $\mathcal Z^k$ decreases rapidly as k increases.

\section{Spatial and momentum density distributions}

Before we investigate the two-particle correlation functions and the chaotic parameter in HBT
interferometry, it is useful to examine the system spatial and momentum density distributions,
$\rho(\textbf{\emph{r}})$ and $\rho(\textbf{\emph{p}})$.  From the equations of the one-body
density matrices in configuration and momentum spaces, (\ref{wfs}), (\ref{G1A}), (\ref{G1Am}),
and (\ref{wfm}), we get
\begin{eqnarray}
\rho(\textbf{\emph{r}})&=&G^{(1)}_{\omega_{\rho}\omega_z}(\textbf{\emph{r}},\textbf{\emph{r}})
=\bigg(\frac{1}{\pi a_{\rho}^2}\bigg)\bigg(\frac{1}{\pi a_z^2}\bigg)^{\!\frac{1}{2}} \exp\bigg(\!\!-
\frac{\brho^2}{a_{\rho}^2}-\frac{z^2}{a_z^2}\bigg) A_{\omega_{\rho}\omega_z}
(\textbf{\emph{r}},\textbf{\emph{r}}),
\end{eqnarray}
\begin{eqnarray}
\rho(\textbf{\emph{p}})=G^{(1)}_{\omega_{\rho}\omega_z}(\textbf{\emph{p}},\textbf{\emph{p}})
=\bigg(\frac{a_{\rho}^2}{\pi\hbar^2}\bigg)\bigg(\frac{a_z^2}{\pi\hbar^2}\bigg)^{\!\frac{1}{2}}
\exp\bigg(\!\!-\frac{\textbf{\emph{p}}_{\rho}^2 a_{\rho}^2}{\hbar^2} -\frac{p_z^2 a_z^2}{\hbar^2}
\bigg) A_{\omega_{\rho}\omega_z} (\textbf{\emph{p}},\textbf{\emph{p}}),
\end{eqnarray}
where
\begin{eqnarray}
A_{\omega_{\rho}\omega_z}(\textbf{\emph{r}},\textbf{\emph{r}})=\frac{\mathcal Z}{1-\mathcal Z}
+\sum_{k=1}^{\infty} \mathcal Z^k \Big[\tilde{g}_0(\textbf{\emph{r}},\textbf{\emph{r}};
\beta k\hbar\,\omega_{\rho},\beta k\hbar\,\omega_z)-1\Big],
\end{eqnarray}
\begin{eqnarray}
A_{\omega_{\rho}\omega_z}(\textbf{\emph{p}},\textbf{\emph{p}})=\frac{\mathcal Z}{1-\mathcal Z}
+\sum_{k=1}^{\infty} \mathcal Z^k \Big[\tilde{g}_0(\textbf{\emph{p}},\textbf{\emph{p}};
\beta k\hbar\,\omega_{\rho},\beta k\hbar\,\omega_z)-1\Big],
\end{eqnarray}
where
\begin{eqnarray}
&&\tilde{g}_0(\textbf{\emph{r}},\textbf{\emph{r}};\tau_{\rho},\tau_z)=\frac{1}{(1-e^{-2\tau_{\rho}})}
\frac{1}{(1-e^{-2\tau_z})^{1/2}}\cr\cr
&&\times \exp\bigg[-\frac{1}{a_{\rho}^2}\frac{\brho^2(\cosh\tau_{\rho}-1 -\sinh\tau_{\rho})}
{\sinh\tau_{\rho}} -\frac{1}{a_z^2}\frac{z^2(\cosh\tau_z-1 -\sinh\tau_z)} {\sinh\tau_z}\bigg],
\end{eqnarray}
\begin{eqnarray}
&&\tilde{g}_0(\textbf{\emph{p}},\textbf{\emph{p}};\tau_{\rho},\tau_z)=\frac{1}{(1-e^{-2\tau_{\rho}})}
\frac{1}{(1-e^{-2\tau_z})^{1/2}}\cr\cr
&&\times \exp\bigg[ -\frac{a_{\rho}^2}{\hbar^2}\frac{\textbf{\emph{p}}_{\rho}^2(\cosh\tau_{\rho}-1
-\sinh\tau_{\rho})}{\sinh\tau_{\rho}} -\frac{a_z^2}{\hbar^2}\frac{p_z^2(\cosh\tau_z-1 -\sinh
\tau_z)}{\sinh\tau_z}\bigg].
\end{eqnarray}

\begin{figure}[tbp]
\includegraphics[width=0.55\columnwidth]{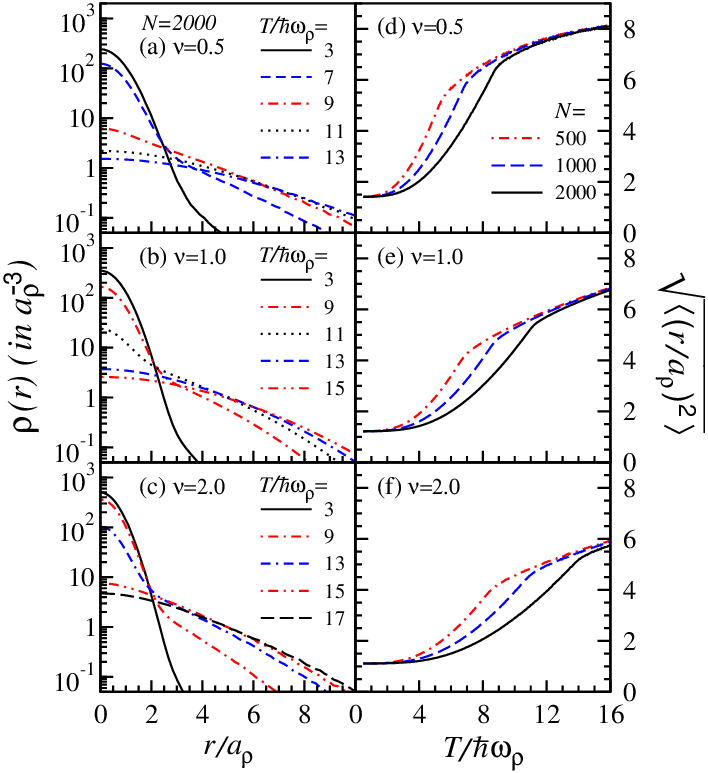}
\caption{(Color online) The spatial density distributions (a)--- (c) and the root-mean-square (d)--- (f)
in unit of $a_{\rho}$ for the systems with $N=2000$, $\nu=$0.5, 1, and 2. }
\label{zrho_r}
\end{figure}

\begin{figure}[tbp]
\includegraphics[width=0.55\columnwidth]{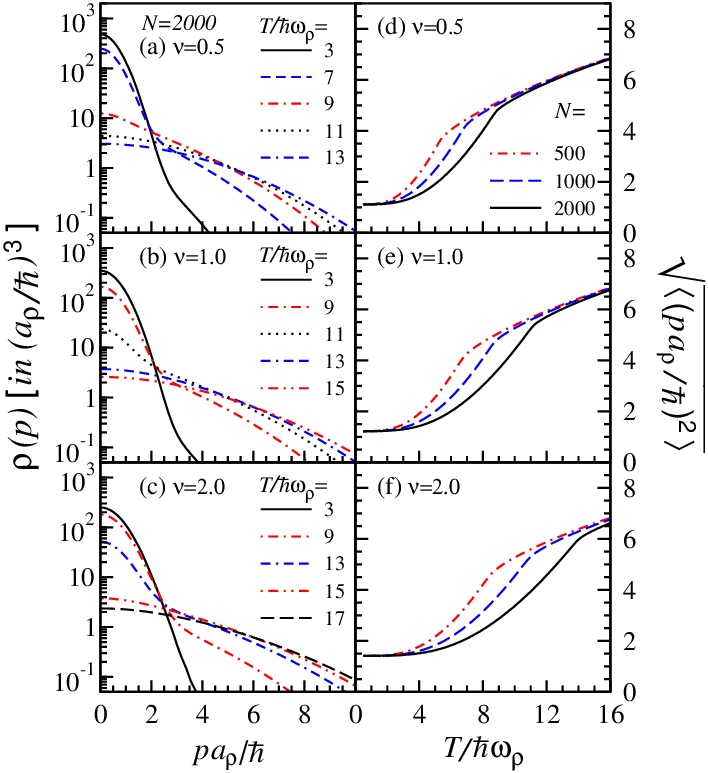}
\caption{(Color online) The momentum density distributions (a)--- (c) and the root-mean-square of momentum
distributions (d)--- (f) in unit of $\hbar/a_{\rho}$ for the systems as the same in Fig. \ref{zrho_r}. }
\label{zrho_p}
\end{figure}

We plot in Figs. \ref{zrho_r}(a)--- \ref{zrho_r}(c) the spatial density distributions as a function of
the dimensionless variable $r/a_{\rho}$ for the systems with $N=2000$, $\nu=$0.5, 1, and 2.  At the low
temperature $T/\hbar\omega_{\rho}=3$, the system distributes in a small spatial region corresponding to
a substantial fraction of condensation.  One can see that the densities at smaller
$r/a_{\rho}$ reduce obviously when temperature increases higher than the critical temperatures, which
are about 9.40, 11.85, and 14.93 $\hbar\,\omega_{\rho}$ for $\nu=$0.5, 1, and 2, respectively.  In Figs.
\ref{zrho_r}(d)--- \ref{zrho_r}(f), we plot the root-mean-square (RMS) of the distributions in unit of
$a_{\rho}$ for the systems with $N=$2000, 1000, and 500.  It can be seen that the RMS increases with
temperature.  For a fixed temperature the RMS decreases with the increasing of the particle number $N$
and the ratio $\nu$, respectively.  It is because that the condensed fraction increases with $N$ and
$\nu$ for a fixed temperature (see Fig. \ref{f0}).  The RMS has a rapid increase in the transition region.
In Fig. \ref{zrho_p} we show the momentum density distributions as a function of the dimensionless variable
$p\,a_{\rho}/\hbar$ and the RMS of the momentum distributions in unit of $\hbar/a_{\rho}$ for the systems
as the same in Fig. \ref{zrho_r}.  One observes that the widths of the momentum distributions for the low
temperature $T/\hbar \omega_{\rho}=3$ are small.  The momentum densities at smaller $p\,a_{\rho}/\hbar$
reduce obviously when temperature increases higher than the critical temperature $T_c$.  The RMS for the
momentum distributions increase with temperature and decrease with the number of particles $N$ as the
spatial densities behaving.  However, the momentum RMS displays different behaviours as a function of
$\nu$ at different temperatures.  At $T\sim0$ (complete condensation), the RMS increases with $\nu$ unlike
that of the RMS for the spatial density distributions.  It is because that the whole phase volume is fixed
for a given $N$ at zero temperature, and the system with smaller $\nu$ has larger spatial volume (larger
$a_z$).  But at the middle temperature $T\sim 8\hbar\,\omega_{\rho}$, the RMS decreases with $\nu$ because
the condensed fraction increases with $\nu$.

\begin{figure}[tbp]
\includegraphics[width=0.55\columnwidth]{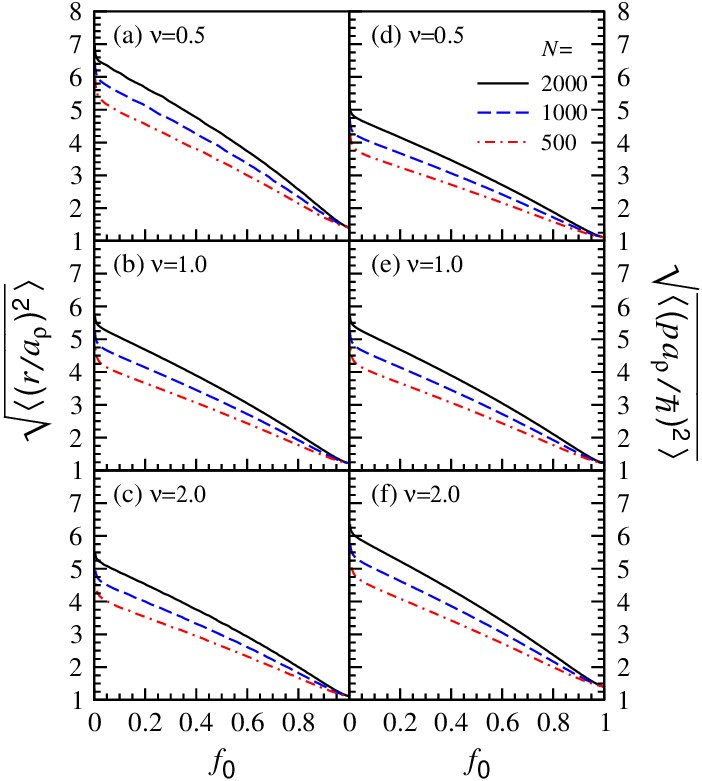}
\caption{(Color online) The root-mean-squares of the distributions of spatial densities (a)--- (c) and
momentum densities (d)--- (f) as a function of $f_0$ for the systems with different $N$ and $\nu$. }
\label{zfrms}
\end{figure}

In Fig. \ref{zfrms}, we plot the results of RMS of the spatial and momentum density distributions as
a function of the condensed fraction $f_0$ for different $N$ and $\nu$ values.  Both the RMS results of
space and momentum decrease with $f_0$ and increase with $N$.  For fixed $f_0$ and $N$, the space RMS
decreases with the ratio $\nu$, and the momentum RMS increases with $\nu$.  It is because that a smaller
$\nu$ corresponds to a larger $a_z$, and thus a larger spatial volume for fixed $a_{\rho}$.  A wider
spatial distribution leads to a narrower momentum distribution for the system with fixed $f_0$ and $N$
(fixed $T$ and $N$).

\section{Chaotic parameter $\lambda$ in HBT interferometry}

\subsection{Evaluation of two-particle correlation functions}

The two-particle correlation function in momentum space is defined as
\begin{equation}
\label{Cp1p2}
C(\textbf{\emph{p}}_{\!1},\textbf{\emph{p}}_2)=\frac{G^{(2)}(\textbf{\emph{p}}_{\!1},{\bf p}_2;
\textbf{\emph{p}}_{\!1},\textbf{\emph{p}}_2)}{G^{(1)}(\textbf{\emph{p}}_{\!1},\textbf{\emph{p}}_{\!1})
G^{(1)}(\textbf{\emph{p}}_2,\textbf{\emph{p}}_2)}.
\end{equation}
From Eqs. (\ref{G1Am}), and (\ref{G2m2}), the two-particle correlation function can be written as
\begin{equation}
\label{Cp1p2A}
C(\textbf{\emph{p}}_{\!1},\textbf{\emph{p}}_2)=1 +\frac{|A_{\omega_{\rho}\omega_z}
(\textbf{\emph{p}}_{\!1},\textbf{\emph{p}}_2)|\,^2 -[\mathcal Z/(1-\mathcal Z)]\,^2} {A_{\omega_{\rho}
\omega_z}(\textbf{\emph{p}}_{\!1},\textbf{\emph{p}}_{\!1}) A_{\omega_{\rho}\omega_z}(\textbf{\emph{p}}_2,
\textbf{\emph{p}}_2)}.
\end{equation}
With the solution $z$ obtained for given $T/\hbar\,\omega_{\rho}$, $N$, and $\nu$ as discussed in
Sec. II and shown in Figs. \ref{zTN} and \ref{zTN2}, we can use Eq. (\ref{A2}) to calculate
$A_{\omega_{\rho}\omega_z}(\textbf{\emph{p}}_i,\textbf{\emph{p}}_{\!j})$ $(i,j=1,2)$ and obtain the
correlation function $C(\textbf{\emph{p}}_{\!1},\textbf{\emph{p}}_2)$ from Eq. (\ref{Cp1p2A}).  At a
very low temperature, the system is completely condensed.  In this case, $A_{\omega_{\rho}\omega_z}
(\textbf{\emph{p}}_{\!1},\textbf{\emph{p}}_2)=\mathcal Z/(1-\mathcal Z)$ and $C(\textbf{\emph{p}}_{\!1},
\textbf{\emph{p}}_2)=1$.  On the other hand, at a very high temperature, the system is completely
uncondensed, we have
\begin{equation}
C(\textbf{\emph{p}}_{\!1},\textbf{\emph{p}}_2)=1 +\frac{|A_{\omega_{\rho}\omega_z}
(\textbf{\emph{p}}_{\!1},\textbf{\emph{p}}_2)|^2} {A_{\omega_{\rho}\omega_z}(\textbf{\emph{p}}_{\!1},
\textbf{\emph{p}}_{\!1}) A_{\omega_{\rho}\omega_z}(\textbf{\emph{p}}_2,\textbf{\emph{p}}_2)}.
\end{equation}

\begin{figure}[tbp]
\includegraphics[width=0.58\columnwidth]{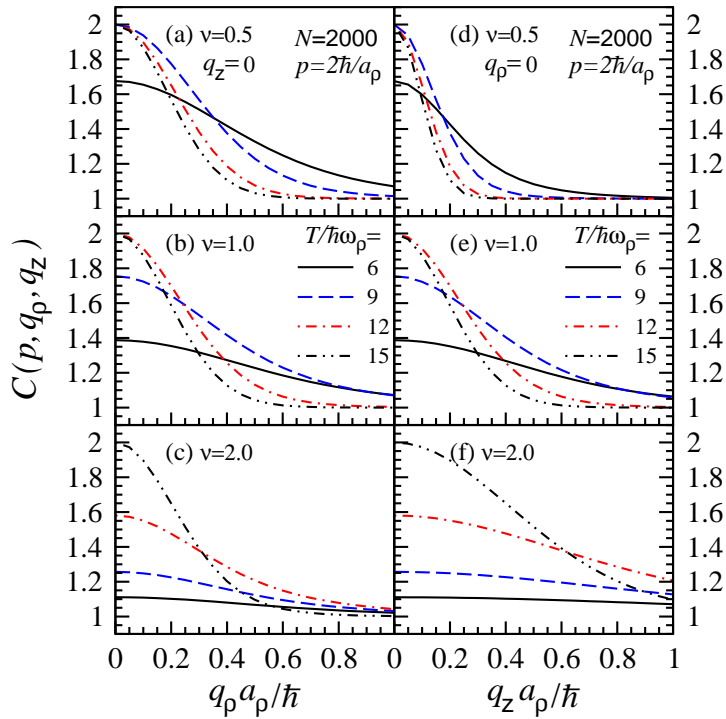}
\caption{(Color online) The two-particle HBT correlation functions $C(p,q_{\rho},q_z=0)$ (a)--- (c)
and $C(p,q_{\rho}=0,q_z)$ for the systems with $N=2000$.  The average moment of the particle pair $p=2
\hbar/a_{\rho}$. } \label{Cqrz}
\end{figure}

In HBT analyses, it is convenient to introduce the average and relative momenta of the particle pair,
$\textbf{\emph{p}}=(\textbf{\emph{p}}_{\!1}+\textbf{\emph{p}}_2)/2$ and $\textbf{\emph{q}}=
\textbf{\emph{p}}_{\!1} -\textbf{\emph{p}}_2$, as variables.  Then, the correlation function
$C(\textbf{\emph{p}},\textbf{\emph{q}})$ can be obtained from $C(\textbf{\emph{p}}_{\!1},
\textbf{\emph{p}}_2)$ by summing over $\textbf{\emph{p}}_{\!1}$ and $\textbf{\emph{p}}_2$ for given
$\textbf{\emph{p}}$ and $\textbf{\emph{q}}$.  In Figs. \ref{Cqrz}(a)--- \ref{Cqrz}(c), we plot the
transverse HBT correlation functions $C(p,q_{\rho},q_z=0)$ for the fixed particle number of system
and the average momentum of the particle pair, $N=2000$ and $p=2\hbar/a_{\rho}$.  The width of the
correlation function decreases with temperature because the spatial distribution of the system increases
with temperature (see Fig. \ref{zrho_r}).  When temperature increases, the condensed fraction $f_0$
decreases, and the intercepts of the correlation functions at $q=0$ increase.  For a fixed temperature,
the intercept decreases with $\nu$ increasing because the condensed fraction increases with $\nu$ (see
Fig. \ref{f0}).  In Figs. \ref{Cqrz}(d)--- \ref{Cqrz}(f), we plot the longitudinal HBT correlation functions
$C(p,q_z, q_{\rho}=0)$ for the systems as the same in Figs. \ref{Cqrz}(a)--- \ref{Cqrz}(c), respectively.
It can be seen that the widths of the correlation functions decrease with temperature.  Because the
longitudinal size of the system, which is related to $a_z$, is larger for a smaller $\nu$ when $a_{\rho}$
is fixed, the widths of the longitudinal correlation functions for the smaller $\nu$ are smaller than those
for the larger $\nu$, respectively.  For $\nu=1$, the widths of the transverse and longitudinal correlation
functions for the same temperature are equaled.  For $\nu>1$ or $\nu<1$, the widths of the longitudinal
correlation functions are smaller or larger than the corresponding widths of the transverse correlation
functions.  The intercepts of the transverse and longitudinal correlation functions at $q=0$ are equaled
for the same temperature and $\nu$ values because the condensed fraction $f_0$ is fixed for given $N$,
$T$, and $\nu$.

\begin{figure}[tbp]
\includegraphics[width=0.57\columnwidth]{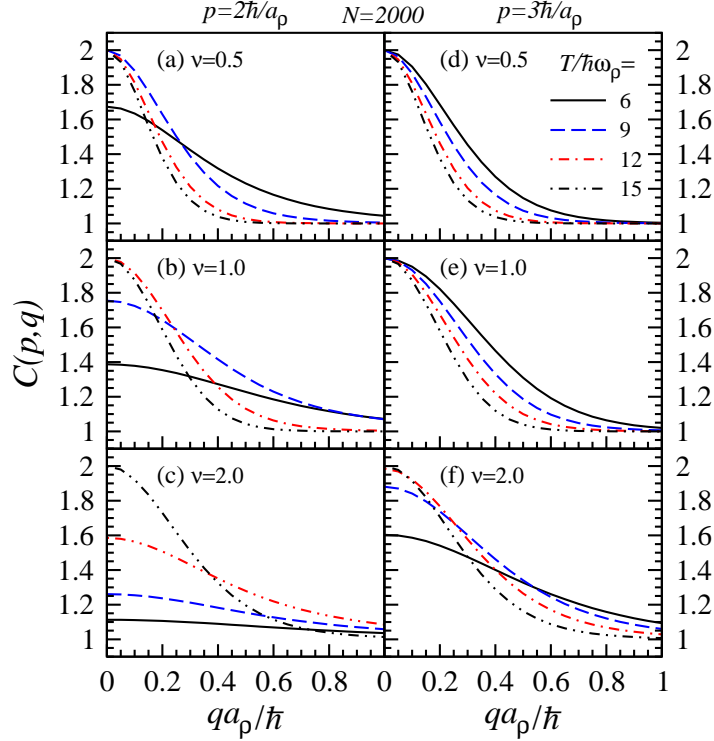}
\caption{(Color online) The two-particle HBT correlation functions $C(p,q)$ for the systems with $N=2000$. }
\label{Cpq2000}
\end{figure}

\begin{figure}[tbp]
\includegraphics[width=0.57\columnwidth]{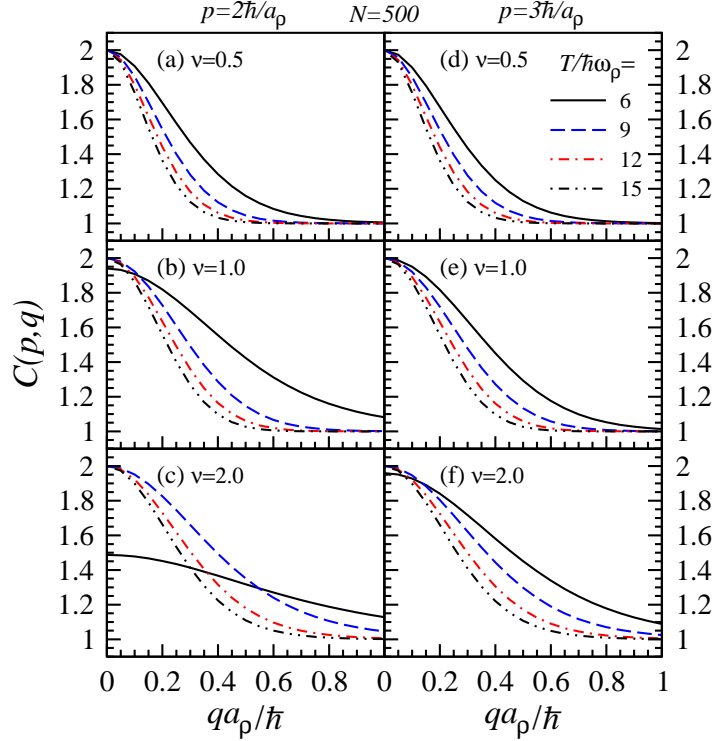}
\caption{(Color online) The two-particle HBT correlation functions $C(p,q)$ for the systems with $N=500$. }
\label{Cpq500}
\end{figure}

In Fig. \ref{Cpq2000}, we plot the two-particle HBT correlation functions, $C(p,q)$, for the systems
with $N=2000$, and the average momenta of the particle pair $p=2$ and $3\hbar/a_{\rho}$.  At the lower
temperatures, the widths of the correlation functions for $p=2\hbar/a_{\rho}$ are larger than the
corresponding results for $p=3\hbar/a_{\rho}$.  And, the intercepts of the correlation functions at
$q=0$ for the lower average momentum of the particle pair are smaller than those for the higher average
momentum.  The reason for these is that the particles with larger momenta are averagely at the uncondensed
high-energy states and thus corresponding to a larger spatial distribution and higher chaotic degree at a
fixed temperature.  The systems with the temperatures much higher than $T_c$ are completely uncondensed.
For the completely uncondensed systems, the widths of the correlation functions for the lower and higher
momenta are almost the same, and the intercepts equal to unity.  The increase of $\lambda$ with momentum
is also predicted in the pion laser model \cite{CsoZim97,ZimCso97}, and the similar behavior may also arise
due to the effect of $\eta'$ decay \cite{Cso10,Ver10}.  In Fig. \ref{Cpq500}, we plot the HBT correlation
functions, $C(p,q)$, for the systems with the particle number $N=500$, and the average momenta of particle
pair $p=2$ and $3\hbar/a_{\rho}$.  Because the condensed fraction decreases with $N$ decreasing, the
intercepts of the correlation functions at $q=0$ for the lower temperatures are higher than the
corresponding results of the intercepts for $N=2000$ as shown in Fig. \ref{Cpq2000}.

\subsection{The effect of Bose-Einstein condensation on the chaotic parameter $\lambda$}

In HBT analyses the chaotic parameter $\lambda$ is introduced phenomenologically to represent the intercept
of the HBT correlation function at zero relative momentum of the particle pair,
\begin{equation}
\lambda\,(\textbf{\emph{p}})=[\,C(\,\textbf{\emph{p}},\textbf{\emph{q}}=0\,)-1\,].
\end{equation}
From Eqs. (\ref{G2m}) and (\ref{Cp1p2}), the chaotic parameter can be expressed by the momentum density
$\rho(\textbf{\emph{p}})=G^{(1)}(\textbf{\emph{p}},\textbf{\emph{p}})$ as
\begin{equation}
\lambda\,(\textbf{\emph{p}})=1-N_0^2 \big[|U_0(\textbf{\emph{p}})|^2/\rho(\textbf{\emph{p}})\big]^2.
\end{equation}

\begin{figure}[tbp]
\includegraphics[width=0.45\columnwidth]{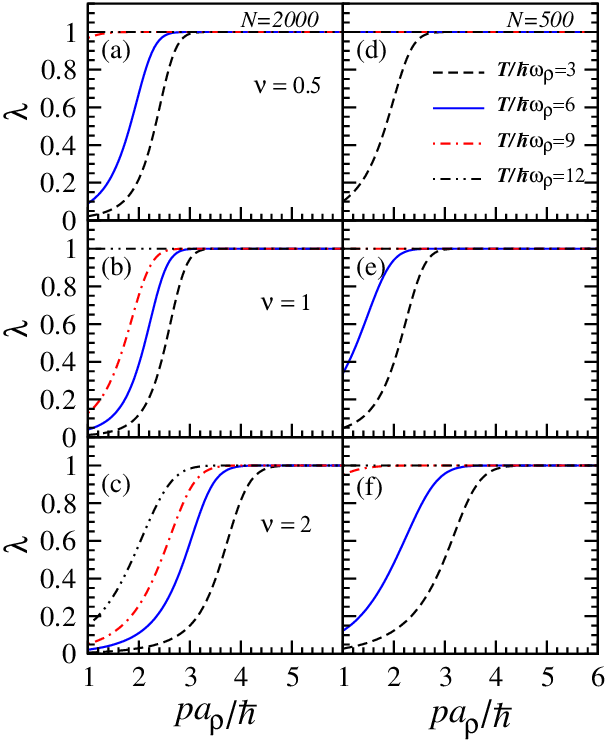}
\caption{(Color online) The values of $\lambda$ as a function of momentum and temperature for the systems
with $N=$2000, $N=$500, $\nu=$ 0.5, 1, and 2, respectively. } \label{lambda}
\end{figure}

\begin{figure}[tbp]
\includegraphics[width=0.45\columnwidth]{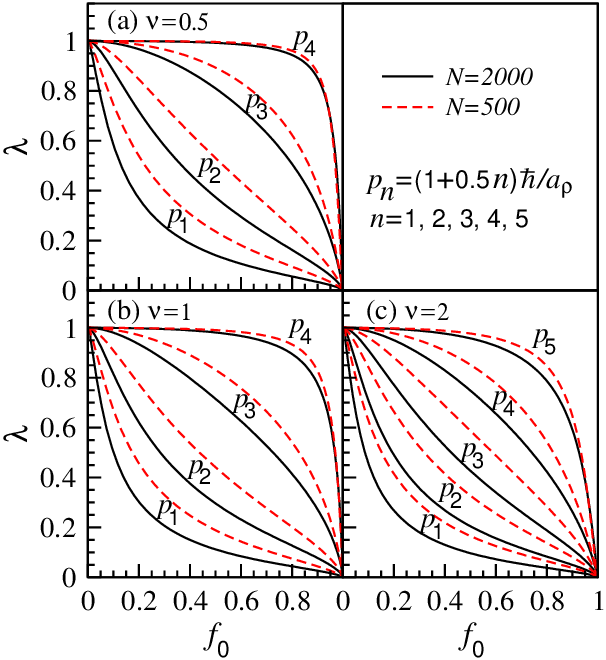}
\caption{(Color online) The values of $\lambda$ as a function of $f_0$ and momentum for the systems with
$N=$ 2000, 500, $\nu=$ 0.5, 1, and 2, respectively. } \label{plambda}
\end{figure}

In Fig. \ref{lambda} we plot the chaotic parameter $\lambda$ as a function of momentum for different
temperatures, $N$, and $\nu$.  It can be seen that the values of $\lambda$ increase with the momentum and
temperature.  For fixed temperature, the values of $\lambda$ for the same momentum decrease with $\nu$ and
$N$.  It is because that the system condensed fraction $f_0$ increases with $\nu$ and $N$ for fixed $T$.
Figure \ref{plambda} (a), (b), and (c) show $\lambda$ as a function of the condensed fraction $f_0$ for
$\nu=$ 0.5, 1, and 2, respectively.  One can see that $\lambda$ decreases with $f_0$.  Because the particles
with larger momenta are averagely at the uncondensed high-energy states, the $\lambda$ values for the larger
momenta are larger.  They decrease with $f_0$ slowly at smaller $f_0$, and drop down to zero only when $f_0
\to 1$.

\subsection{$\lambda$ values in high energy heavy ion collisions}

At the final stage of high energy heavy ion collisions, pions will be scattered out and the source will
be in freeze-out state.  The number of identical pions is about several hundreds or thousands at RHIC
or LHC energy, and the range of the freeze-out temperature $T_f$ is about 80 --- 165 MeV.  Because the
temperature is of the order of the pion rest mass, a relativistic treatment of the pion motion is needed.
The eigenvalue equation for the relativistic pion with only a scalar interaction as in Eq. (\ref{Vrz})
is \cite{Won07}
\begin{equation}
\bigg[\frac{\textbf{\emph{p}}^2}{2m}+V(\textbf{\emph{r}})\bigg]U(\textbf{\emph{r}})=\frac{E^2-m^2}{2m}
U(\textbf{\emph{r}})\equiv \epsilon U(\textbf{\emph{r}}).
\end{equation}
The eigenenergy of the relativistic pion is
\begin{equation}
E_n=\sqrt{m^2+2m\epsilon_n},~~~~~~n=0,1,2,\cdots,
\end{equation}
where $\epsilon_n$ is given by Eq. (\ref{enen}) and the corresponding eigenfunction is
\begin{equation}
U_n(\textbf{\emph{r}})=U_{n_x}(x,\omega_x)U_{n_y}(y,\omega_y)U_{n_z}(z,\omega_z),
\end{equation}
where the one-dimension wave function $U_{n_x}(x,\omega_x)$ is given by Eq. (\ref{wfx}) and $\omega_x=
\omega_y =\omega_{\rho}$.  We introduce $\tilde E_n$ to measure the relative energy levels to the
ground-state energy,
\begin{equation}
\tilde E_n=E_n-\sqrt{m^2+2m\hbar\,(\omega_{\rho}+\frac{1}{2}\omega_z)}.
\end{equation}
The number of identical bosons of the system in relativistic case is then
\begin{equation}
\label{Nrel}
N=N_0+N_T=\frac{\mathcal Z}{1-\mathcal Z}+\sum_{n>0}^{\infty}\frac{g_n \mathcal Z\,e^{-\beta\tilde E_n}}
{1-\mathcal Z\,e^{-\beta\tilde E_n}},
\end{equation}
and the densities of space and momentum are
\begin{equation}
\label{rhor_rel}
\rho(\textbf{\emph{r}})=G^{(1)}(\textbf{\emph{r}},\textbf{\emph{r}})=\sum_{n=0}^{\infty}\frac{g_n
\mathcal Z\,e^{-\beta\tilde E_n}}{1-\mathcal Z\,e^{-\beta\tilde E_n}}|U_n(\textbf{\emph{r}})|^2,
\end{equation}
\begin{equation}
\label{rhop_rel}
\rho(\textbf{\emph{p}})=G^{(1)}(\textbf{\emph{p}},\textbf{\emph{p}})=\sum_{n=0}^{\infty}\frac{g_n
\mathcal Z\,e^{-\beta\tilde E_n}}{1-\mathcal Z\,e^{-\beta\tilde E_n}}|U_n(\textbf{\emph{p}})|^2,
\end{equation}
where $U_n(\textbf{\emph{p}})$ is the eigenfunctions in momentum space.

\begin{figure}[tbp]
\includegraphics[width=0.5\columnwidth]{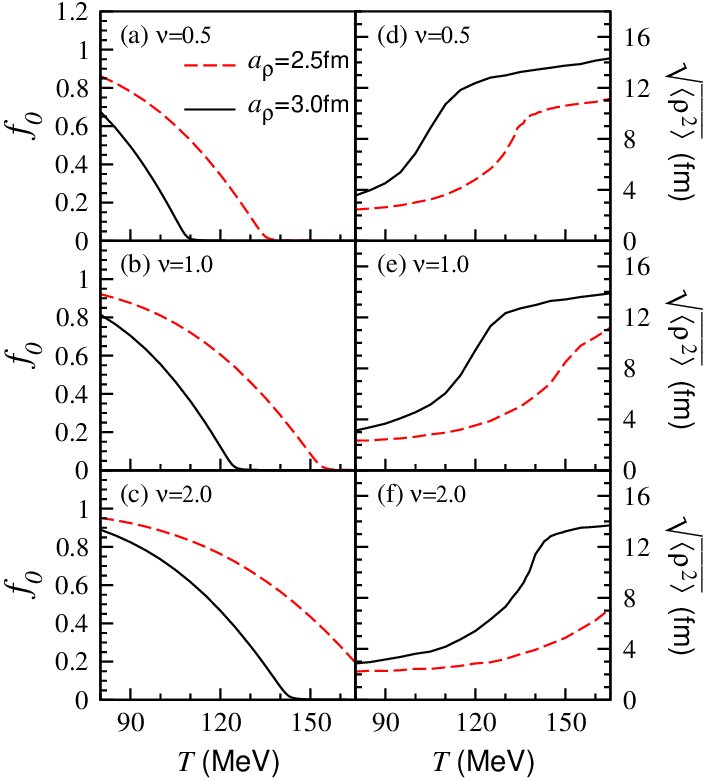}
\caption{(Color online) The condensed fraction $f_0$ and the RMS of transverse coordinate
$\sqrt{\langle \rho^2 \rangle}$ as a function of temperature for the systems with $N=$ 2000, $a_{\rho}=$
2.5 and 3.0 fm. } \label{f0_rel}
\end{figure}

\begin{figure}[tbp]
\includegraphics[width=0.42\columnwidth]{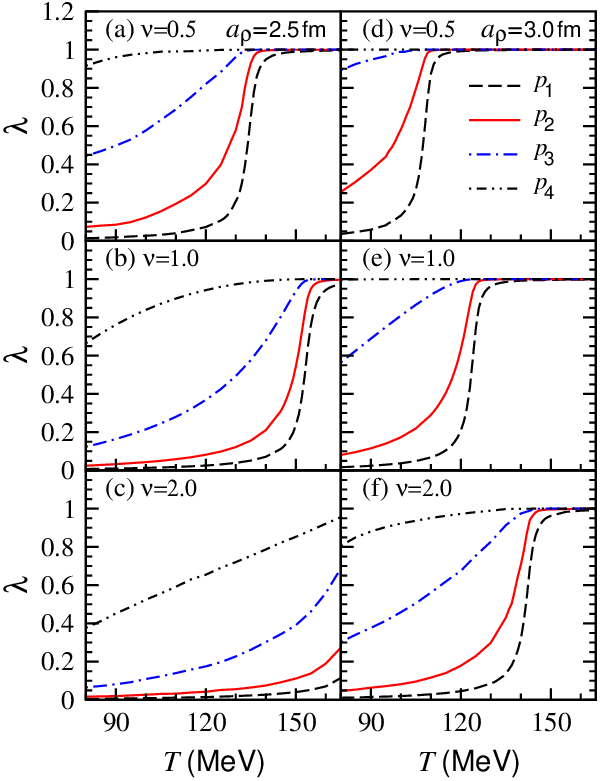}
\caption{(Color online) The values of $\lambda$ as a function of momentum and temperature for the systems
with $N=$ 2000, $a_{\rho}=$ 2.5 and 3.0 fm.  $p_n=50n$ MeV/$c$ ($n=1,2,3,4$). } \label{lambda_rel}
\end{figure}

In the calculations for relativistic pion, $m=140$ MeV is fixed, and we need input $a_{\rho}$ (or
$\omega_{\rho}$) and $a_z$ (or $\nu$).  From Eq. (\ref{Nrel}) we can obtain the fugacity parameter
$\mathcal Z$ for fixed $N$ and $T$ numerically. Then, we can obtain $\rho(\textbf{\emph{r}})$ and
$\rho(\textbf{\emph{p}})$ from Eqs. (\ref{rhor_rel}) and (\ref{rhop_rel}).

In Fig. \ref{f0_rel} we plot the condensed fraction $f_0$ and the RMS of transverse coordinate
$\sqrt{\langle \rho^2 \rangle}$ in the temperature range 80 --- 165 MeV and for the systems with $N=$
2000, $a_{\rho}=$ 2.5 and 3.0 fm.  It can be seen that $f_0$ decreases from about 1 to zero in this
temperature range.  Correspondingly, $\sqrt{\langle \rho^2 \rangle}$ values increase with temperature
and have obvious enhancements in the transition region from the condensed phase to uncondensed phase.
We observe that the effect of condensation is sensitive to the parameter $a_{\rho}$.  For the smaller
$a_{\rho}$, the values of $f_0$ are larger and the values of $\sqrt{\langle \rho^2 \rangle}$ are smaller
as compared to the results for the larger $a_{\rho}$.  For fixed $a_{\rho}$ and $T$, $f_0$ increases and
$\sqrt{\langle \rho^2 \rangle}$ decreases with $\nu$.  In Fig. \ref{lambda_rel} we plot the $\lambda$
values in the temperature range 80 --- 165 MeV for the systems with $N=$ 2000, $a_{\rho}=$ 2.5 and 3.0
fm.  Here the momenta $p_n=50n$ MeV/$c$ ($n=1,2,3,4$).  For the smaller momenta, the values of $\lambda$
have rapid increases in the transition region from the condensed phase to uncondensed phase.  However,
the values of $\lambda$ for the larger momenta are higher even at the lower temperatures.  It is because
that most of the particles with large momenta are from the uncondensed high-energy states.  The values
of $\lambda$ increase with $\nu$ and decrease with $a_{\rho}$ because of $f_0$ decreases with $\nu$ and
increases with $a_{\rho}$.

\begin{figure}[tbp]
\includegraphics[width=0.5\columnwidth]{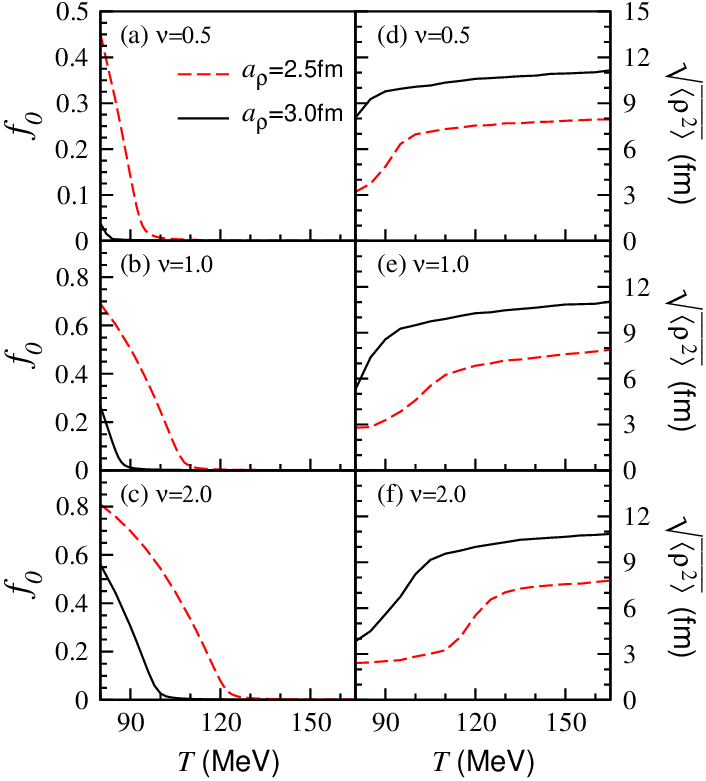}
\caption{(Color online) The condensed fraction $f_0$ and the RMS of transverse coordinate
$\sqrt{\langle \rho^2 \rangle}$ as a function of temperature for the systems with $N=$ 500, $a_{\rho}=$
2.5 and 3.0 fm. } \label{f0_rel5}
\end{figure}

\begin{figure}[tbp]
\includegraphics[width=0.42\columnwidth]{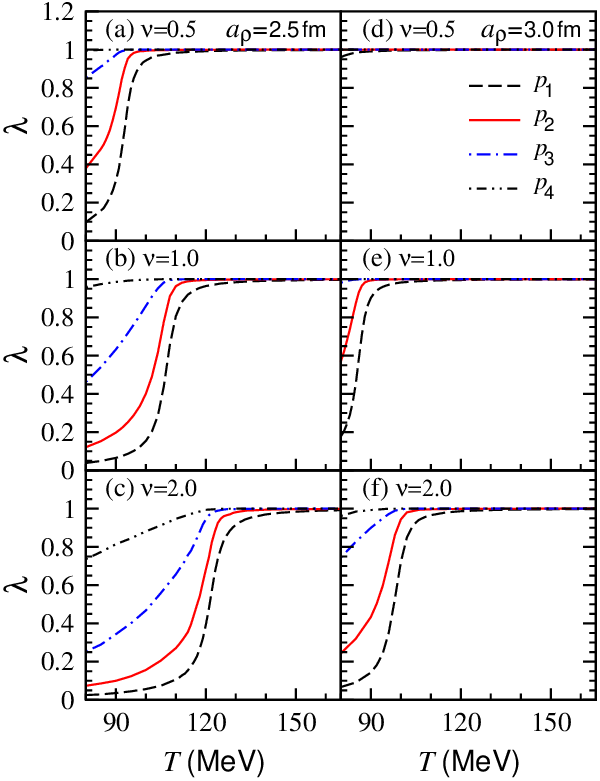}
\caption{(Color online) The values of $\lambda$ as a function of momentum and temperature for the systems
with $N=$ 500, $a_{\rho}=$ 2.5 and 3.0 fm.  $p_n=50n$ MeV/$c$ ($n=1,2,3,4$). } \label{lambda_rel5}
\end{figure}

In Figs. \ref{f0_rel5} and \ref{lambda_rel5} we plot the quantities $f_0$, $\sqrt{\langle\rho^2\rangle}$,
and $\lambda$ for the systems with $N=$500 in the temperature range 80 --- 165 MeV.  Because the critical
temperature for $N=$500 is much smaller than that for $N=$2000, the values of $f_0$ are smaller and about
zero at most of the temperatures.  Correspondingly, the RMS values have obvious increases in only lower
temperature regions and increase slowly at most of the temperatures.  As compared to the results for the
systems with $N=$2000, the values of $\lambda$ for $N=$500 are larger.  For $a_{\rho}=$3.0 and $\nu=$0.5,
the values of $\lambda$ are unity in almost the whole temperature range because the system is completely
in the uncondensed phase at almost the whole temperatures.

\begin{figure}[tbp]
\includegraphics[width=0.42\columnwidth]{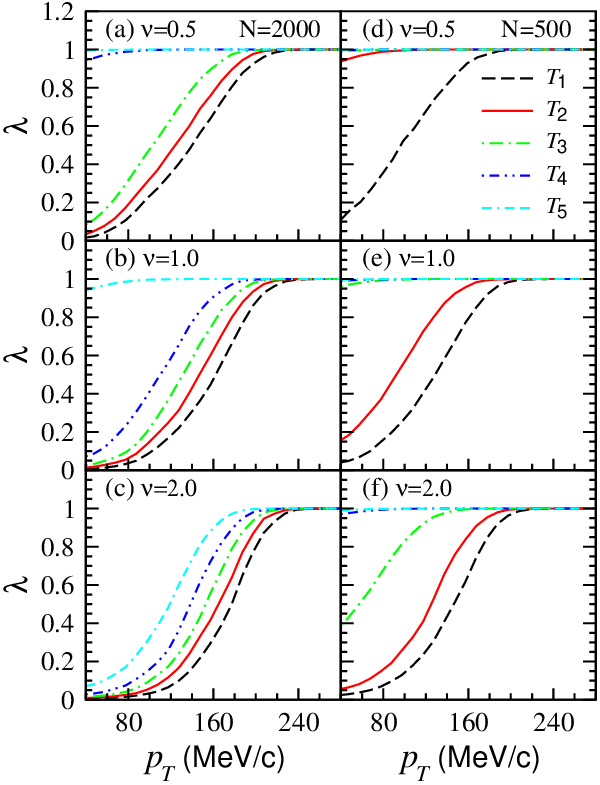}
\caption{(Color online) The values of $\lambda$ as a function of transverse momentum and temperature for
the systems with $a_{\rho}=$ 2.5 fm, $N=$ 2000 and 500.  $T_n=80+20(n-1)$ MeV $(n=1,2,3,4,5). $} \label{lambda_relpt}
\end{figure}

In Fig. \ref{lambda_relpt} we further plot $\lambda$ as a function of pion transverse momentum $p_T$
for the systems with $a_{\rho} =2.5$ fm, $N=$ 2000 and 500, and $\nu=$ 0.5, 1.0, and 2.0, respectively.
Here the temperature $T_n=80+20(n-1)$ MeV $(n=1,2,3,4,5)$.  It can be seen that the values of $\lambda$
increase with $p_T$ and temperature.  For fixed $p_T$ and $T$, the values of $\lambda$ decrease with
increasing particle number $N$ and frequency ratio $\nu$.

In the heavy ion collisions at RHIC, the identical pion multiplicity is about several hundreds.  The
calculations indicate that in this case the effect of Bose-Einstein condensation on the chaotic parameter
$\lambda$ in two-pion HBT interferometry may be negligible.  However, the identical pion multiplicity in
the heavy ion collisions at LHC energy may reach to several thousands.  In this case the effect of
Bose-Einstein condensation on the two-pion HBT measurements of $\lambda$ may be considerable and should
be taken into account.  From Fig. \ref{f0_rel} we observe that the average values of $\sqrt{\langle \rho^2
\rangle}$ in the transition region are about 6.5 fm for $a_{\rho}=2.5$ fm and 8.5 fm for $a_{\rho}=3.0$
fm.  The recent two-pion interferometry measurements at LHC indicate that the values of the transverse HBT
radius $R_{\rm side}$ are in the range of 4 --- 7 fm \cite{ALI11}.  Considering the transverse expansion
of the actual particle-emitting source may decrease the transverse HBT radii from the RMS of the source
transverse coordinate distribution \cite{Wie99,Her95,Cha95,Yin12}, these average results of $\sqrt{\langle
\rho^2 \rangle}$ are in a reasonable range.  Further investigating the effect of Bose-Einstein condensation
on the HBT measurements of source radii and chaotic degree in high energy heavy ion collisions, based on a
more realistic model of evolving source, will be of great interest.

In our calculations the particle number of system, $N$, is fixed.  In this case, the $\lambda$ value for
an event ensemble with the multiplicity distribution, $P_N$, is
\begin{eqnarray}
\label{lambda1}
\lambda'=\frac{\sum_N P_N \lambda_N}{\sum_N P_N},
\end{eqnarray}
where $\lambda_N$ is the $\lambda$ value for the system with the fixed $N$, as calculated in Sec. IV,
\begin{equation}
\lambda_N=\frac{G^{(2)}_N(\textbf{\emph{p}},\textbf{\emph{p}}; \textbf{\emph{p}},\textbf{\emph{p}})} {G^{(1)}_N(\textbf{\emph{p}},\textbf{\emph{p}}) G^{(1)}_N(\textbf{\emph{p}},\textbf{\emph{p}})}-1.
\end{equation}
Another calculation of the ensemble $\lambda$ value is
\begin{equation}
\label{lambda2}
\lambda''=\frac{\sum_N P_N\, G^{(2)}_N(\textbf{\emph{p}},{\bf p}; \textbf{\emph{p}},
\textbf{\emph{p}})}{\sum_N P_N\, G^{(1)}_N(\textbf{\emph{p}},\textbf{\emph{p}})
G^{(1)}_N(\textbf{\emph{p}},\textbf{\emph{p}})}-1.
\end{equation}
Because multiplicity is a observable in high energy heavy ion collisions, and the event number for
a certain multiplicity is large and in principle unstinted (by prolonging experimental time), the
measurement of the ``differential" chaotic parameter $\lambda_N$ is viable and useful in probing
the particle source coherence.  On the other hand, the appearance of the condensation may modify
identical pion multiplicity distribution in high energy heavy ion collisions \cite{CsoZim97,ZimCso97},
and thus affect the ``integrated" chaotic parameter values $\lambda'$ and $\lambda''$.  Investigating
the influences of Bose-Einstein condensation on $\lambda'$ and $\lambda''$ will be of interest.

\section{Summary and discussion}

We examine the spatial and momentum distributions, two-particle HBT correlation functions, and the
chaotic parameter $\lambda$ in HBT interferometry for the systems of boson gas within the harmonic
oscillator potentials with anisotropic frequencies in transverse and longitudinal directions.  The
effects of system temperature, particle number, and the average momentum of the particle pair on the
chaotic parameter are investigated.  Because of Bose-Einstein condensation of boson gas the system is
highly condensed at low temperature.  This leads to the narrower spatial and momentum distributions
of the system, larger width of the HBT correlation functions, and smaller $\lambda$ values.  When
temperature increasing the system becomes uncondensed gradually.  The values of $\lambda$ increase
with temperature and rapidly reach to unity when temperature is closed to and higher than the critical
value $T_c$.  For fixed particle number and temperature of system, the $\lambda$ values increase with
the average momentum of the particle pair because the particles with large momenta are averagely at
the uncondensed high-energy states.  The results of $\lambda$ are sensitive to the ratio, $\nu=\omega_z
/\omega_{\rho}$, of the frequencies in longitudinal and transverse directions.  They are smaller for
larger $\nu$ when $\omega_{\rho}$ is fixed.  Because the critical temperature of the system increases
with the particle number of system, $N$, the effect of Bose-Einstein condensation on the interferometry
measurements of $\lambda$ will be significant for a large $N$ system.  In the heavy ion collisions at
LHC energy the identical pion multiplicity may reach to several thousands.  In this case the system
may possible reach to a considerable condensation.  The effect of the condensation on the chaotic
parameter in two-pion interferometry is worth considering in earnest.  Because we did not consider
the particle charge in the model, the results are only suitable for neutral particles, for example
$\pi^0$.  The investigation of the possible change of the results for neutral and charged particles
will be of interest.

In experimental HBT analyses, the correlation function is obtained by the ratio of the correlated two
particle momentum distribution ${\rm Cor}(\textbf{\emph{q}},\textbf{\emph{p}})$ to the uncorrelated
two particle momentum distribution (background) ${\rm Uncor}(\textbf{\emph{q}}, \textbf{\emph{p}})$ \cite{Zajc84,WNZ93}.  Here ${\rm Cor}(\textbf{\emph{q}},\textbf{\emph{p}})$ is constructed by the
identical particle pairs in which the two particles are from the same event.  And ${\rm Uncor}
(\textbf{\emph{q}},\textbf{\emph{p}})$ is constructed by the two identical particles from the different
events with the same global conditions (cuts), such as within a certain region of particle multiplicity
(centrality), rapidity, or pseudorapidity, and so on.  For the different events, the degree of
Bose-Einstein condensation may be different although they have the same global conditions.  The
difference of the condensation degree for the events may affect the correlated and uncorrelated
momentum distributions, and then affect the measurement of the chaotic parameter, corresponding to
Eq. (\ref{lambda2}).  However, this influence may become smaller and smaller in principle by imposing
stricter and stricter global condition cuts, and it will be an interesting research issue.  On the
other hand, the model we used in the calculations is only a static system.  To make the problem
tractable we used the mean-field of harmonic oscillator potential.  In fact, the particle-emitting
sources produced in high energy heavy ion collisions are expanded and the interactions between the
particles in the source are complicated.  Further investigating the effect of Bose-Einstein
condensation on the HBT measurements of source radii and chaotic degree in high energy heavy ion
collisions, based on a more realistic model of evolving source, will be of great interest.

\begin{acknowledgments}
This research was supported by the National Natural Science Foundation of China under Grant Nos. 11075027
and 11275037.
\end{acknowledgments}


\begin{thebibliography}{99}

\bibitem{Gyu79}
M. Gyulassy, K. K. Kauffman, and L. W. Wilson, Phys. Rev. C 20 (1997) 2267.

\bibitem{Won94}
Cheuk-Yin Wong, {\it Introduction to High-Energy Heavy-Ion Collisions}, World Scientific Publishing
Company, Singapore, 1994, Chap. 17.

\bibitem{Wie99}
U. Wiedemann, U. Heinz, Phys. Rep. 319 (1999) 145.

\bibitem{Wei00}
R. M. Weiner,  Phys. Rept. 327 (2000) 249.

\bibitem{Lis05}
M. A. Lisa, S. Pratt, R. Soltz, and U. Wiedemann, Ann. Rev. Nucl. Part. Sci. 55 (2005) 357.

\bibitem{CsoZim97}
T. Cs\"{o}rg\H{o} and J. Zim\'{a}nyi, Phys. Rev. Lett. 80 (1986) 916.

\bibitem{ZimCso97}
J. Zim\'{a}nyi and T. Cs\"{o}rg\H{o}, Acta Phys. Hung. New Ser. Heavy Ion Phys. 9 (1999) 241.

\bibitem{Won07}
Cheuk-Yin Wong and Wei-Ning Zhang, Phys. Rev. C 76 (2007) 034905.

\bibitem{Pol96}
H. D. Politzer, Phys. Rev. A 54 (1996) 5048.

\bibitem{Nar99}
M. Naraschewski and R. J. Glauber, Phys. Rev. A 59 (1999) 4595.

\bibitem{Via06}
J. Viana Gomes, A. Perrin, M. Schellekens, D. Boiron, C. I. Westbrook, and M. Belsley, Phys.
Rev. A 74 (2006) 053607.

\bibitem{Cso10}
T. Cs\"{o}rg\H{o}, R. V\'ertesi, and J. Sziklai, Phys. Rev. Lett. 105 (2010) 182301.

\bibitem{Ver10}
R. V\'ertesi, T. Cs\"{o}rg\H{o}, and J. Sziklai, Phys. Rev. C 83 (2011) 054903.

\bibitem{ALI11}
K. Aamodt et al. (ALICE Collaboration), Phys. Lett. B 696 (2011) 328.

\bibitem{Her95}
M. Herrmann and G. F. Bertsch, Phys. Rev. C 51 (1995) 328.

\bibitem{Cha95}
S. Chapman, P. Scotto, and U. Heinz, Phys. Rev. Lett. 74 (1995) 4400.

\bibitem{Yin12}
H. J. Yin, J. Yang, and W. N. Zhang, Phys. Rev. C 86 (2012) 024914.

\bibitem{Zajc84}
W. A. Zajc et al. Phys. Rev. C 29 (1984) 2173.

\bibitem{WNZ93}
W. N. Zhang, Y. M. Liu, S. Wang, Q. J. Liu, J. Jiang, D. Keane, Y. Shao, S. Y. Chu, S. Y. Feng,
Phys. Rev. C 47 (1993) 795.

\end{thebibliography}
\end{document}